**Femtosecond induced transparency and absorption in the extreme ultraviolet by coherent coupling of the He 2s2p ($^1P^o$) and 2p$^2$ ($^1S^e$) double excitation states with 800 nm light**

Zhi-Heng Loh,[1] Chris H. Greene,[2] and Stephen R. Leone[1*]

[1] Departments of Chemistry and Physics, University of California, Berkeley, CA 94720, USA, and Chemical Sciences Division, Lawrence Berkeley National Laboratory, Berkeley, CA 94720, USA

[2] Department of Physics and JILA, University of Colorado, Boulder, CO 80309, USA

**Abstract**

Femtosecond high-order harmonic transient absorption spectroscopy is used to observe electromagnetically induced transparency-like behavior as well as induced absorption in the extreme ultraviolet by laser dressing of the He 2s2p ($^1P^o$) and 2p$^2$ ($^1S^e$) double excitation states with an intense 800 nm field. Probing in the vicinity of the 1s$^2$ → 2s2p transition at 60.15 eV reveals the formation of an Autler-Townes doublet due to coherent coupling of the double excitation states. Qualitative agreement with the experimental spectra is obtained only when optical field ionization of both double excitation states into the $N = 2$ continuum is included in the theoretical model. Because the Fano $q$-parameter of the unperturbed probe transition is finite, the laser-dressed He atom exhibits both enhanced transparency and absorption at negative and positive probe energy detunings, respectively.

[*] Corresponding author. Electronic address: srl@berkeley.edu

1. Introduction

The interaction of light with a coherent atomic ensemble leads to important quantum interference phenomena, such as coherent population trapping [1], electromagnetically induced transparency (EIT) [2], and laser-induced continuum structure [3]. The dominant physics underlying the phenomena listed above involves the use of two laser fields to dress the eigenstates of a three-level system (Fig. 1), which is most commonly presented in either a lambda or ladder configuration. In coherent population trapping, two laser fields with similar Rabi frequencies ($\Omega_1 \sim \Omega_2$) are employed to manipulate the state populations of individual atoms or molecules, giving rise to many applications in high-resolution spectroscopy [1]. In EIT, a strong coupling laser field modifies the linear and nonlinear susceptibility of the optical medium, which is then probed by a weak laser field ($\Omega_1 \gg \Omega_2$) [2]. Replacement of the discrete eigenstate at the vertex of a lambda system by a continuum allows a dressing laser to embed a discrete atomic state into the otherwise featureless continuum; the resulting laser-induced continuum structure is then observed by a weak probe field [3]. Altogether, these processes can be used to manipulate the linear and nonlinear susceptibilities of matter from the near-infrared to ultraviolet, resulting in applications such as slowed and stored light, quantum information processing, and enhanced nonlinear frequency conversion [4]. Similarly exciting applications are found in the realm of chemistry, such as implementation of quantum optics on single molecules [5], selective population of dressed states and manipulation of the quantum mechanical phase of atoms [6], all-optical alignment of nonpolar molecules [7], and determination of the transition dipole moment between arbitrary pairs of molecular states [8].

Presently there is great interest in the application of light in the x-ray regime, produced by high-order harmonics, free electron lasers, and sliced synchrotron sources, to investigate novel coherent x-ray optical phenomena [9]. This work reports the observation of EIT-like behavior in the extreme ultraviolet (XUV) by coherent coupling of two adjacent double excitation states in He, probing with laser-produced high-order harmonics. While previous demonstrations of EIT reveal a dramatic increase in transparency on a sub-linewidth scale, and are further accompanied by phenomena such as slowed light and enhanced nonlinear optical susceptibilities [2], the EIT-like phenomenon observed in this work is characterized solely by an increase in transmission over the entire unperturbed lineshape. Existing experimental studies on the interaction of electric fields with the He double excitation states are confined to measuring the effect of weak (10 V/cm) to moderate ($10^5$ V/cm) static (dc) electric fields on the photoabsorption [10] and fluorescence spectra [11,12]. These studies show the effect of dc Stark mixing on the photoabsorption cross-section and fluorescence/autoionization branching ratios of these states, as well as to reveal states that cannot be observed in the field-free photoabsorption spectra. Theoretical studies on the effect of ac fields on He double excitation states predict the transformation of the original Fano lineshape [13] into an Autler-Townes doublet [14] due to Rabi flopping between a pair of coherently coupled double excitation states [15,16]. While these theoretical studies on the double excitation states of He have yet to be verified experimentally, a study employing two-photon absorption probing in the ultraviolet demonstrated autoionization suppression via coherent population trapping in Mg [17].

In a new apparatus, constructed to investigate ultrafast chemical dynamics by transient absorption on core level transitions, a femtosecond pump laser pulse is used to initiate an ultrafast photophysical or photochemical transformation, which is then probed after a variable time delay by an XUV pulse. The XUV probe pulse is produced on a table-top, laser-based setup by high-order harmonic generation [18,19]. In an initial study, the spin-orbit and hole orbital alignment population distributions were determined for $Xe^+$ ions produced by strong-field ionization of Xe atoms [20]. To further explore the basic applicability of this apparatus on isolated atoms, we employ it here to observe the coherent coupling of the $2s2p$ ($^1P^o$) and $2p^2$ ($^1S^e$) double excitation states of He by a strong 800 nm field ($F_L^{max} = 1 \times 10^8$ V/cm). The $2s2p$ and $2p^2$ states are located at 60.15 [21] and 62.06 eV [22] above the $1s^2$ ground state, respectively (Fig. 2). The experimental results are compared to theoretical spectra calculated following the formalism developed in Ref. [23] for treating the effect of strong laser fields on autoionizing states.

**2. Experimental methods**

The schematic of the experimental setup is illustrated in Fig. 3. Briefly, the amplified output from a commercial Ti:sapphire laser system (2.4 W, 800 nm, 42 fs, 1 kHz) is sent to a 20:80 beamsplitter to produce the optical dressing and high-order harmonic generation (HHG) beam, respectively. High-order harmonics in the XUV region are generated by focusing the laser light into a 7 cm long, 150 $\mu$m internal diameter capillary filled with $1.1 \times 10^4$ Pa of Ne [24]. The estimated photon flux at the source is $10^5$ photons per pulse for the high-order harmonic centered at 60.2 eV. A pair of 0.2 $\mu$m thick Al foils is used to reject the residual 800 nm light and transmit the high-order harmonics. After reflection by a toroidal mirror, the high-order harmonics are refocused into a 2 mm long gas cell filled with $1.3 \times 10^4$ Pa of He. The transmitted XUV radiation is spectrally dispersed in a home-built spectrometer and detected with a thermoelectrically cooled CCD camera. A dielectric-coated mirror with a 1 mm diameter internal bore hole allows the optical dressing beam to overlap with the XUV probe beam in a collinear geometry. Previous work on optical strong-field ionization of Xe gives an estimate of 30 fs FWHM for the XUV pulse duration [20]. Scanning knife-edge measurements provide a beam waist of 21 $\mu$m for the high-order harmonics. The spectrometer resolution is determined to be 0.18 eV FWHM from the observed linewidth of the $1s^2 \rightarrow 2s2p$ transition. Note that the spectrometer measures the transmission spectrum directly (instead of the absorption spectrum). The transmission spectrum therefore needs to be convolved with the spectrometer resolution function before computing the absorption spectrum. For high sample densities, this procedure can yield absorption lineshapes that seem to have anomalously small Fano $q$-parameters. (The Fano $q$-parameter is a measure of the relative ratio of the transition to the continuum from the ground state compared to the transition via an autoionizing state.) However, with the spectrometer resolution as the only adjustable parameter, the static absorption spectrum obtained from our experiments can be fit to a Fano lineshape using a literature value for the $q$-parameter.

The dressing pulse energy incident on the gas cell is 0.10 mJ, the beam waist at the focus is 62 $\mu$m, and the pulse duration is 42 fs FWHM. The temporal profile of the dressing pulse measured using spectral interferometry for direct electric-field reconstruction [25] after the He gas cell is identical to that obtained for an evacuated gas cell. These parameters yield a peak intensity (electric field) of $1.4 \times 10^{13}$ W/cm$^2$ ($1.0 \times 10^8$ V/cm) for the dressing pulse. Transient absorption spectra are obtained by using spectra collected at –500 fs time delay as the reference (a constant background offset due to stray pump beam light incident on the CCD camera precludes the use of the pump-off spectrum as the reference spectrum); a negative time delay implies that the probe pulse arrives at the sample before the pump pulse. Each spectrum is obtained from the average of 320 sets of data and error bars correspond to 95% confidence interval limits.

## 3. Results and discussion

The transient absorption spectra acquired at different dressing-probe time delays are shown in Fig. 4. Since the probe pulse envelope acts as a windowing function to select a range of dressing field strengths experienced by the atom, a change of the dressing-probe time delay allows the atom to be probed at various electric field strengths of the dressing laser. Such an approach is reasonable only when the duration of the XUV probe pulse is shorter than or comparable to that of the dressing laser pulse, a condition that occurs when the XUV radiation is produced via high-order harmonic generation [20,26]. The transient absorption spectrum at a time delay of 0 fs is a mirror image of the static absorption spectrum (Fig. 4 inset). This suggests that the absorption spectrum of the laser-dressed atom at high electric fields is essentially featureless (the unknown fraction of dressed atoms in the interaction volume prevents the photoabsorption spectrum of the dressed atom from being retrieved experimentally), thereby allowing the reduction of the static absorption to dominate the transient absorption spectrum. Decreasing the field strength of the dressing laser by increasing the time delay leads to a simultaneous weakening and narrowing of the absorption bleach at 60.1 eV, while a positive ΔOD feature gradually develops on the low energy side. At the same time, an absorption shoulder at 60.5 eV becomes apparent. The similarity of the ΔOD values at positive XUV detunings and dressing-probe time delays of 40 – 80 fs is due to the presence of a post-pulse on the 800 nm dressing laser; the post-pulse appears 88 fs after the main peak with a maximum intensity of 13% that of the main peak.

To elucidate the origin of the spectral features observed in the transient absorption spectra, theoretical ΔOD spectra are calculated following the formalism developed in Ref. [23]. In the limit of a weak probe field such that the probe Rabi frequency $\Omega_{ga}(t,r) = \frac{1}{2}\mu_{ga}F_X(t,r) \ll 1$, the instantaneous photoabsorption rate as a function of time and radial coordinate is given by [27]

$$R(t,r) = \gamma_g(t,r) + \Omega_{ga}^2(t,r) \frac{\left(1-\frac{1}{q^2}\right)\left(\delta_T^2\Gamma_a + \frac{1}{4}\Gamma_a\Gamma_b^2 + \Gamma_b\Omega_{ab}^2(t,r)\right) + \frac{1}{q}\left(4\delta_T^2\delta_X - 4\delta_T\Omega_{ab}^2(t,r) + \delta_X\Gamma_b^2\right)}{\left(\delta_T\delta_X - \frac{1}{4}\Gamma_a\Gamma_b - \Omega_{ab}^2(t,r)\right)^2 + \frac{1}{4}(\delta_T\Gamma_a + \delta_X\Gamma_b)^2}, \quad (1)$$

where the indices g, a, and b are used to denote the $1s^2$ ground, $2s2p$, and $2p^2$ states, respectively. In the above equation, $\mu_{ij}$ is the dipole matrix element for the $i \rightarrow j$ transition, $\gamma_g(t,r)$ is the laser-induced width of the ground state, q is the Fano q-parameter for the probe transition ($q = -2.75$) [21], $E_i^r$ and $\Gamma_i$ ($i = a,b$) are resonance energies and widths of the double excitation states, respectively ($\Gamma_a = 1.37 \times 10^{-3}$ a.u., $\Gamma_b = 2.16 \times 10^{-4}$ a.u.) [21,28], $\Omega_{ab}(t,r) = \frac{1}{2}\mu_{ab}F_L(t,r)$ is the Rabi frequency for the $2s2p \rightarrow 2p^2$ transition, $F_X(t,r)$ and $F_L(t,r)$ are the electric field envelopes of the XUV probe and 800 nm dressing laser, respectively, $\delta_T = E_L + E_X - E_b^r$ is the total photon energy detuning, $\delta_X = E_X - E_a^r$ is the XUV probe detuning, and $E_L$ and $E_X$ are the 800 nm and XUV photon energies, respectively. $\gamma_g(t,r)$ is calculated from an expression given in Ref. [16]. An eigenchannel R-matrix calculation [29] yields a value of 2.17 a.u. for the dipole matrix element $\mu_{ab}$. Note that Eq. (1), which describes the formation of an Autler-Townes doublet, reduces to the expression for EIT [30] in the limit of a negligible interference from the direct photoionization channel, i.e., when $q \gg 1$ — the underlying physical mechanisms for Autler-Townes doublet formation and EIT are the same. The theoretical transient absorption spectra, obtained after spatial- and temporal averaging to account for the finite spatio-temporal extent of the dressing and probe laser pulses, are convolved with a Gaussian broadening of 0.18 eV FWHM to account for the spectrometer resolution. The resultant theoretical spectra are shown in Fig. 5 (left panel).

The theoretical ΔOD spectra (Fig. 5a) show that in addition to the experimentally observed negative and positive features at probe detunings of –0.1 eV and 0.1 eV, respectively, a second set of absorption peaks at probe detunings of –0.4 eV and 0.7 eV for dressing-probe time delays of 0 fs and 20 fs would also occur. These are not observed experimentally. An increase in the dressing-probe time delay causes the latter set of peaks to move inward. The theoretical photoabsorption spectra of the dressed He atom prior to convolution with the spectrometer resolution function (Fig. 5b) shows that the second set of transient absorption peaks originate from the Autler-Townes doublet. The integrated photoabsorption cross-section is independent of the dressing-probe time delay, as required by the oscillator sum rule. Increasing the dressing-probe time delay decreases the dressing field strength and hence the Rabi frequency, thereby allowing both members of the Autler-Townes doublet to approach one another. Their deviation from the symmetric double-peaked structure, as seen in conventional double resonance spectra, can be attributed to several factors. First, the use of an ultrashort dressing laser pulse exposes the atom to a range of electric field strengths over the duration of the XUV probe pulse, which leads to a spread in the measured Rabi frequencies $\Omega_{ab}(t,r)$ when probing with a pulse of finite temporal width. This effect is particularly pronounced in the photoabsorption spectrum calculated for a time delay of 20 fs due to the large range of electric field strengths occurring along the slope of the dressing laser temporal profile. Second, the asymmetry of the field-free Fano lineshape and the non-zero dressing laser detuning ($\delta_L = E_L - E_b^r + E_a^r = -0.36$ eV) give rise to components of the Autler-Townes doublet with different peak heights.

Even though the simulation is able to reproduce several experimentally observed spectral features, the peaks predicted at probe detunings of –0.4 eV and 0.7 eV for dressing-probe time delays of 0 fs and 20 fs are noticeably absent in the experimental spectra, suggesting that the theoretical model is incomplete. As noted above, the experimental transient absorption spectra suggest that the absorption spectrum of the laser-dressed atom at high field strengths is featureless. Such an absorption spectrum can be obtained in the limit whereby laser-induced ionization results in severe linewidth broadening. Previous work showed that the laser-induced ionization of the He double excitation states to the $N = 1$ continuum is negligible due to lack of orbital overlap between initial and final configurations [16]. However, for the peak electric field strength of 0.02 a.u. employed in this experiment, optical field ionization of the $2s2p$ and $2p^2$ double excitation states into the $N = 2$ continuum needs to be taken into consideration. In fact, the ionization threshold for the $N = 2$ continuum, located at 65.40 eV above the $1s^2$ ground state [21], is sufficiently close to the resonance energies of the $2s2p$ and $2p^2$ states for over-the-barrier ionization [31] to occur before the peak of the dressing field; the critical field strengths for over-the-barrier ionization of the $2s2p$ and $2p^2$ states into the $N = 2$ continuum are $9.4 \times 10^{-3}$ and $3.8 \times 10^{-3}$ a.u., respectively. At lower field strengths, ionization via a tunneling mechanism is possible. (Tunneling ionization rates into the $N = 3$ continuum, which is located at 72.96 eV above the ground state, are at least 7 orders of magnitude lower compared to those for the $N = 2$ continuum, and can therefore be neglected.) To approximate the contribution of this additional ionization channel, the field-free linewidths in Eq. 1 are replaced with an effective linewidth $\Gamma_{i,eff}(t,r) = \Gamma_i + W_i(t,r)$ ($i = a,b$), where $W_i(t,r)$ is the optical field ionization rate that is dependent on the instantaneous electric field strength of the dressing laser. In the absence of accurate theoretical values, $W_i(t,r)$ is approximated here as $W_i(t,r) = W_{ADK,i}(t,r)$ for $W_i(t,r) < 2/\pi n_i^3$ and $W_i(t,r) = 2/\pi n_i^3$ otherwise, where $W_{ADK,i}(t,r)$ is the Ammosov-Delone-Krainov ionization rate [32], $n_i = (2I_{p,i})^{-1/2}$ is the effective principal quantum number, and $I_{p,i}$ is the ionization potential. (Note that in the high field limit, the optical field ionization rate is not allowed to exceed a maximum value that corresponds to the theoretical maximum expected for a Rydberg level autoionizing state.) Following these corrections to account for optical field ionization of the double excitation states directly into the $N = 2$ continuum, a new set of transient absorption spectra are calculated and the results are shown in Figs. 5c and 5d.

Compared to the previous set of theoretical spectra, significantly better agreement with the experimental data is obtained with the inclusion of optical field ionization into the $N = 2$ continuum (Fig. 5c). Given the simplistic approximation for the field ionization rate, as well as the reduction of the manifold of autoionizing states to include only the $2s2p$ and $2p^2$ states, the agreement between the theoretical and experimental spectra is satisfactory. In spite of the large dressing field detuning, the observed agreement between theoretical and experimental spectra suggests that the dynamic Stark shift due to states outside the three-level subspace is negligible. Inspection of the theoretical photoabsorption spectra of the laser-dressed atom (Fig. 5d) shows that at short dressing-probe time delays, the inclusion of field ionization results in severe linewidth broadening such that the Autler-Townes doublet is barely visible. At a dressing-probe time delay of 0 fs, inclusion of optical field ionization leads to a linewidth broadening of the low energy component of the Autler-Townes doublet from 0.07 to 0.2 eV FWHM. The theoretical photoabsorption spectra also confirm that with decreasing dressing field strengths, the experimentally observed (1) narrowing and weakening of the bleach at 60.1 eV, (2) the development of a positive absorption feature to the low energy side of the bleach, and (3) the formation of a shoulder at 60.5 eV are all spectral signatures of the formation of an Autler-Townes doublet due to coherent coupling of the $2s2p$ and $2p^2$ double excitation states of He.

Under our experimental conditions for the gas density and interaction length, the theoretical photoabsorption spectra obtained in the absence of spectral resolution broadening (Fig. 5d) suggest that the transmission at the peak of the Fano lineshape should increase from <0.1% to 59% upon laser dressing at 0 fs dressing-probe time delay, corresponding to a decrease in optical depth from >6.9 to 0.53. This is an EIT-like phenomenon in the XUV. Note that the enhanced transparency on the sub-linewidth scale exhibited in conventional EIT requires the dressing field to be either on- or near-resonance ($\delta_L \sim \Gamma_a$), and the Rabi frequency of the dressing field to be larger than the linewidths of the autoionizing states ($|\Omega_{ab}|^2 \gg \Gamma_a\Gamma_b$) [4]. Both conditions are not satisfied in the present work due to the large dressing laser detuning and optical field ionization-induced widths $\Gamma_{i,\mathit{eff}}$. In spite of the large dressing laser detuning, however, the transient absorption spectrum at 0 fs time delay reveals that increased transparency over the entire $\delta_X < 0$ region can be achieved due to the large laser-induced linewidth, in agreement with the spectra obtained from theoretical modeling (Figs. 5c and 5d). In the present work, the narrow field-free linewidth (37 meV) compared to the spectrometer resolution (0.18 eV FWHM) obscures the dramatic change in the transparency after the differential transmission spectrum is convolved with the spectrometer resolution function; at 0 fs time delay and $\delta_X = -0.1$ eV, the experimentally observed transmission increases from $19\pm1\%$ to $24\pm1\%$. The choice of another autoionizing resonance with a broader linewidth should allow a more direct experimental observation of EIT at these short wavelengths. Recent theoretical work suggests that EIT in the x-ray should be observed via coherent coupling of the $1s^{-1}3s$ and $1s^{-1}3p$ states of Ne [33]; the Auger decay width of these core-excited states is 0.27 eV.

In the case of He, the window portion of the Fano lineshape gives rise to an effect that is the exact opposite of EIT — electromagnetically induced absorption (EIA). Previous work showed that coherent population trapping involving the quasi-degenerate ground state hyperfine levels of atomic Rb can lead to either enhanced transparency or absorption, depending on the choice of the probe transition [34]. In our experiments, EIA is observed simultaneously with EIT and manifests itself as the positive ΔOD feature at positive probe detunings in the transient absorption spectra. At 0 fs time delay and $\delta_X = 0.1$ eV, for example, the measured transmission decreases from $69 \pm 2\%$ to $52 \pm 2\%$. Note that this EIA effect will be particularly dramatic whenever the XUV light probes at high resolution a Fano minimum whose unperturbed photoabsorption cross section goes to zero. Every autoionizing resonance having $|q| < \infty$ is known to possess one true zero provided it can only decay into a single continuum, like the $2s2p$ $^1P^o$ state probed in this study.

## 4. Concluding remarks

Femtosecond XUV absorption spectroscopy employing a laser-based, high-order harmonic generation setup is used to observe EIT-like behavior in the vicinity of the $1s^2 \to 2s2p$ transition of He at 60.15 eV. This EIT-like behavior arises from the formation of an Autler-Townes doublet induced by coherent coupling of the $2s2p$ ($^1P^o$) and $2p^2$ ($^1S^e$) double excitation states by an 800 nm dressing laser. Theoretical modeling following a formalism for treating the effect of ac fields on autoionizing states reveals that, at the high peak electric fields employed in this study, inclusion of optical field ionization of both double excitation states into the $N = 2$ continuum is needed to reproduce the experimental transient absorption spectra. Even though the large laser-induced ionization widths ($\Gamma_{a,eff}\Gamma_{b,eff} >> |\Omega_{ab}|^2$) and dressing laser detuning ($\delta_L >> \Gamma_a$) are in contradiction to the requirements for conventional EIT, the observation of enhanced transparency over the entire unperturbed lineshape in this work points to the emergence of optical field ionization as an important physical mechanism for achieving an EIT-like phenomenon in the strong field regime. Along with the increase in transparency at negative probe detunings, the finite Fano $q$-parameter of the probe transition allows EIA to exist simultaneously at the window portion of the Fano lineshape at positive probe detunings.

The observation of spectral signatures that are indicative of both EIT and EIA in this work paves the way for further exploration of the interaction of short-wavelength light with coherent media, especially as future free electron lasers and synchrotron slicing light source facilities become available. Applications envisioned include ultrafast optical switching and spectral phase shaping in the XUV. Ultrafast optical switching based on EIT in the x-ray regime has been proposed as a route to generating ultrashort pulses from synchrotron facilities for x-ray pump, x-ray probe experiments [33]. Similarly, spectral phase shaping at these short wavelengths can be used to access coherent control [35] of inner-shell electron dynamics of atoms and molecules [36].

**Acknowledgements**

We are grateful to R. Santra and C. Buth for drawing our attention to the significance of the laser-induced ionization channel, and we thank T. Pfeifer, M. Khalil, and R. E. Correa for useful discussions. C.H.G. was partly supported by the Miller Institute for Basic Research in Science, and partly by DOE (DE-FG02-94ER14413). This work was supported by the NSF ERC for Extreme Ultraviolet Science and Technology (EEC-0310717) and the LDRD program at LBNL, with additional equipment support from DOE (DE-AC02-05CH11351).

**List of figure captions**

Fig. 1. (a) Lambda and (b) ladder coupling schemes for coherent population trapping and EIT. (c) Coupling scheme for laser-induced continuum structure. $\Omega_1$ and $\Omega_2$ represent the Rabi frequencies of the two laser fields.

Fig. 2. Schematic illustration showing the energy levels of the $1s^2$ ground state and the $2s2p$ and $2p^2$ double excitation states. The $N = 1$ and $N = 2$ continua are also shown, along with values of their respective ionization thresholds. The hatched area below the field-free $N = 2$ continuum shows the region that is accessible by optical field ionization.

Fig. 3. Schematic illustration of the experimental setup. The light and dark lines correspond to the 800 nm dressing and XUV probe beam paths, respectively.

Fig. 4. Experimental transient absorption spectra collected at various dressing-probe time delays. ΔOD denotes the change in optical density (absorbance). The static (field-free) absorption spectrum for the $1s^2 \rightarrow 2s2p$ transition is shown in the inset. The solid line corresponds to a spectrally broadened Fano lineshape with resonance energy and $q$-parameter of 60.15 eV and –2.75, respectively.

Fig. 5. Normalized theoretical (a) transient absorption spectra and (b) photoabsorption spectra of the dressed atom calculated following Ref. [23]. The corresponding spectra calculated with the inclusion of optical field ionization of the $2s2p$ and $2p^2$ double

excitation states into the $N = 2$ continuum are shown in (c) and (d). The photoabsorption spectra, which do not include broadening from the finite spectrometer resolution, reveal the formation of the Autler-Townes doublet in the vicinity of the $1s^2 \rightarrow 2s2p$ transition with laser dressing. The transient absorption spectra (a) exhibit peaks at probe detunings of –0.4 eV and 0.7 eV that are not observed experimentally, suggesting that the theoretical model is incomplete. Spectra (c) show that these peaks vanish upon the inclusion of optical field ionization directly into the $N = 2$ continuum. The experimental data points are superposed on the theoretical spectra. The photoabsorption spectra (c) and (d) reveal a dramatic decrease in the peak absorption with laser dressing compared to the static (zero-field) spectrum. This decrease in photoabsorption cross-section leads to EIT-like behavior in the XUV.

**Figure 1**

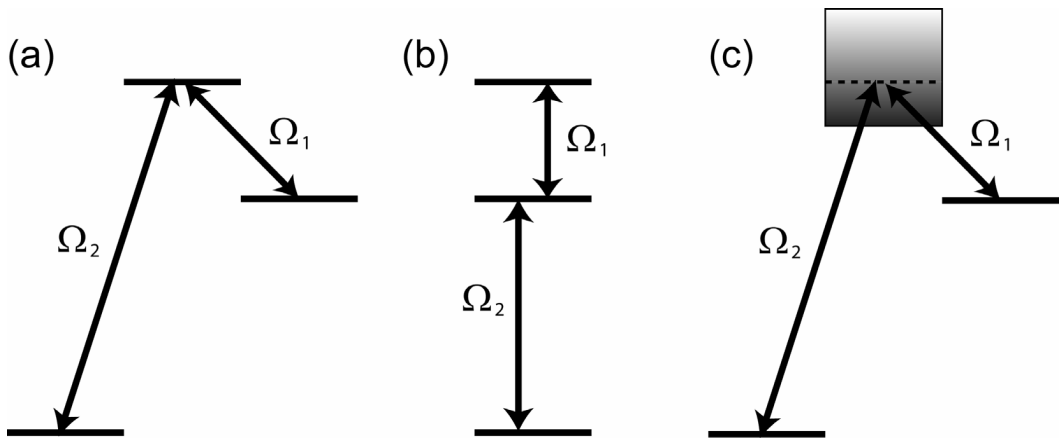

**Figure 2**

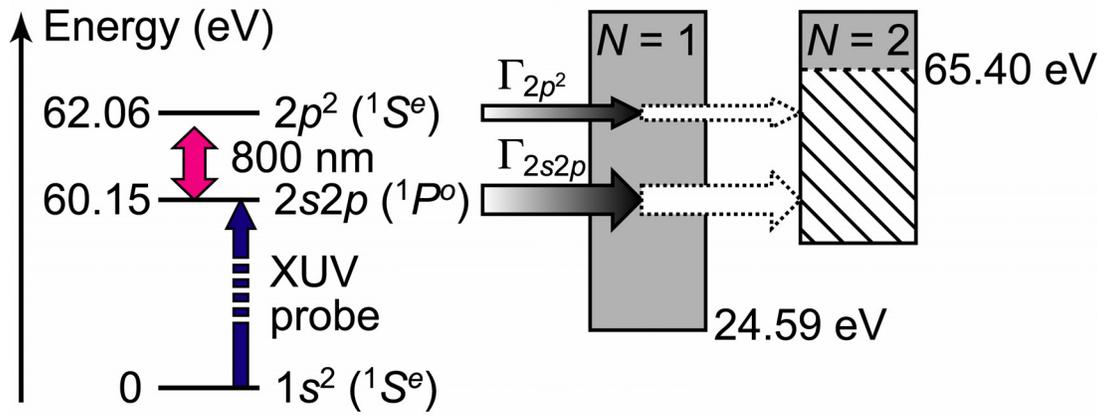

**Figure 3**

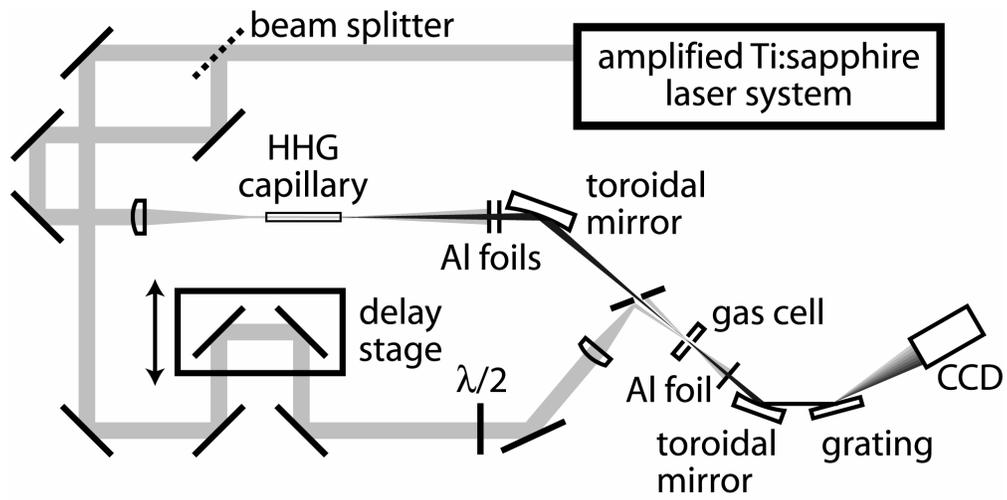

**Figure 4**

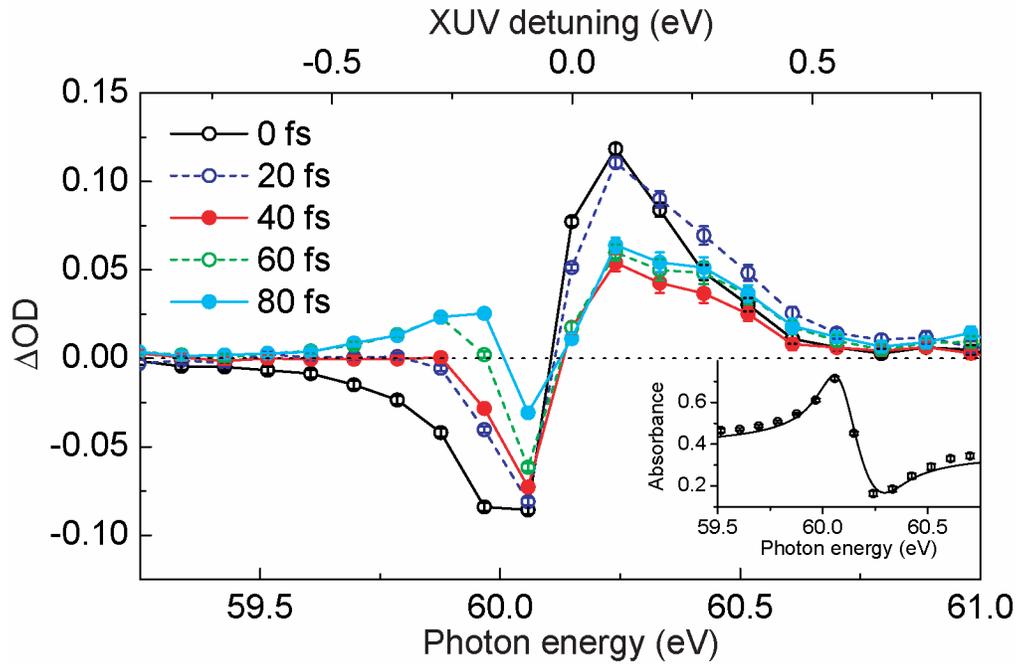

**Figure 5**

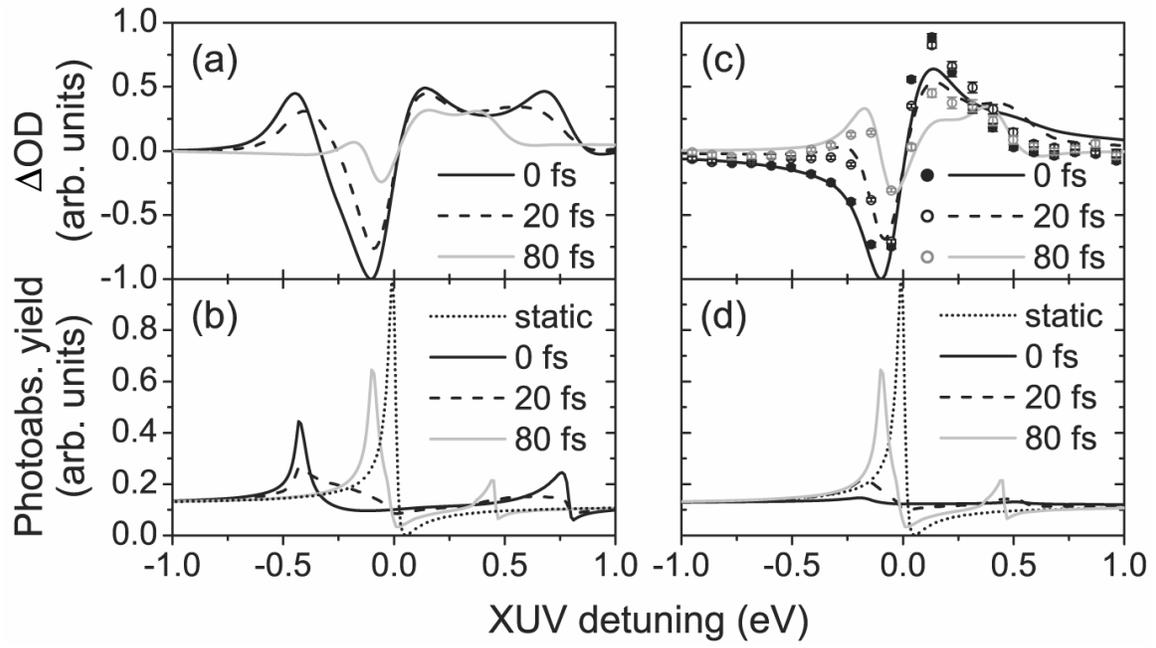